\documentclass[aps,prl,reprint,showpacs,showkeys,groupedaddress,nofootinbib]{revtex4-1}

\usepackage{hyperref}
\usepackage{amsmath}
\usepackage{setspace}
\usepackage{psfrag}
\usepackage{subfigure}
\usepackage{mathtools}
\usepackage{amssymb}
\usepackage{verbatim}
\usepackage{color}
\usepackage{textcomp}
\usepackage[toc,page]{appendix}
\usepackage{graphicx}
\usepackage{bm}
\usepackage{dcolumn}
\usepackage{color}

\newcommand{\be}{\begin{equation}}
\newcommand{\ee}{\end{equation}}
\newcommand{\ben}{\begin{equation*}}
\newcommand{\een}{\end{equation*}}
\newcommand{\bea}{\begin{eqnarray}}
\newcommand{\eea}{\end{eqnarray}}
\newcommand{\bean}{\begin{eqnarray*}}
\newcommand{\eean}{\end{eqnarray*}}
\newcommand{\ti}{\textit}
\newcommand{\tb}{\textbf}
\newcommand{\tu}{\underline}
\newcommand{\na}{\mbox{\boldmath$\nabla$}}
\begin{document}

\title{Effects of the Quantum Potential on lower dimensional models of analogue gravity}

\author{Supratik Sarkar}
\thanks{supratiks@students.iiserpune.ac.in}
\author{Arijit Bhattacharyay}
\thanks{a.bhattacharyay@iiserpune.ac.in}

\affiliation{Indian Institute of Science Education and Research, Pune 411008, India}

\date{December 9, 2015}

\begin{abstract}
We address the issues related to the presence of the quantum potential term in a BEC on the observable Analogue Gravity systems. We show that the quantum potential term apparently gives rise to massive scalar excitations of length scales of the order $\mathcal{O}(1/\xi)$ in the \ti{lower dimensional space}. Since, in 'analogue models' , there is a window for experimental observations generally in $(2+1)$ or even lower dimensional space, one has to take proper account of the presence of these massive excitations to interpret the results.  
\end{abstract}



\maketitle

\section{1. Introduction}
It is possible to create large curvatures of effective space-time as seen by sonic excitations in some condensed matter systems at a very low temperature. Bose-Einstein condensate (BEC) is one of the prominent candidates of all these systems \cite{Barcelo2013}. 
This fact opens up the scope of experimenting on various aspects of quantum fields on curved space-time, e.g., possibility of observing Hawking radiation, cosmological particle production etc. Unruh's seminal work \cite{Unruh1981} practically opened up this field of research which is actively pursued over last couple of decades and a host of theoretical proposals are around \cite{Jacobson1991,Barcelo2011}. Particularly in the context of emergent gravity in BEC, Parentani and coworkers have proposed a number of novel ideas based on density correlations, studying the hydrodynamics over several length scales, surface gravity independent temperature etc \cite{Macher2009,Finazzi2011a,Finazzi2011b,Finazzi2012}. Novel use of density-density correlations inside and outside the horizon has also been exploited by Balbinot \ti{et al} \cite{Balbinot2008}. 
\par
\textcolor{black}{The scattering of sound-wave perturbations from vortex excitations has been investigated by numerical integration of the associated Klein- Gordon equation in order to extract a sizeable fraction of the vortex energy through a mechanism of superradiance by Federici \ti{et al} \cite{Federici2006,Cherubini2005} .}
\par
BEC is a superfluid quantum phase of matter which is considered to be the most probable condensed matter candidate to create analogue gravitational scenarios at a very low (nano Kelvin) temperatures \cite{Garay2000,Garay2001} . In this paper, we're showing that the Lorentz breaking quantum potential term of BEC gives rise to massive excitations.
\par
The small amplitude collective excitations of a uniform density moving phase of BEC obeys the quantum hydrodynamics which, ignoring the quantum potential term \cite{Pitaevskii2003}, can be cast into the equation of a massless free scalar field on a Lorentzian manifold. There has been efforts to regularise the dynamics taking into account the quantum potential term \cite{Fleurov2012}. In 2005, Visser \ti{et al.} have shown the emergence of a massive Klein-Gordon equation considering a two component BEC where a laser induced transition between the two components is exploited \cite{Visser2005}. Liberati \ti{et al.} proposed a weak $U(1)$ symmetry breaking of the analogue BEC model by the introduction of an extra quadratic term in the Hamiltonian to make the scalar field massive \cite{Liberati2009}. Considering the flow in a Laval nozzle, Cuyubamba has shown the emergence of a massive scalar field in the context of analogue gravity arguing for the possibility of observation of quasi-normal ringing of the massive scalar field within the laboratory setup \cite{Cuyubamba2013}.
\par
The healing length$(\xi)$ of a BEC, in the standard condensed matter context, is considered to be $\lambda_{C}/\sqrt{2}$ where $\lambda_{C}$ is the effective Compton wavelength of a particle where the velocity of light is replaced by the velocity of sound. Below this length scale, the quantum potential term of BEC becomes non-negligible and there happens a Lorentz breakdown in the analogue picture. Dispersion becomes important at small length scales. One takes advantage of this fact of knowing the dispersion relation and tries to address the analogue trans-planckian problem in this regime \cite{Unruh1995,Unruh2005,Schutzhold2008,Brout1995,Corley1996} . The basic idea of most of such works is to understand the robustness of the Hawking radiation (Planckian spectrum) in the presence of Lorentz-breaking dispersions.
\par
In the present paper, we are addressing the effects of the presence of the Lorentz-breaking term(i.e. the Quantum Potential term) using a completely different approach through \ti{multiple scale perturbations}.
\par
Our main result in this paper is to show that as a consequence of quantum potential induced Lorentz symmetry breaking of the massless scalar field at smallest length scales, there emerges a massive scalar field at larger length scales on a spacetime of \ti{lower} dimensions. This is a general result within the scope of analogue systems, because it does not take into account anything special about the condensate, namely \ti{multicomponent, special geometry} or \ti{forced symmetry breaking}. 
\par
Present results assume importance in view of the fact that, in actual experimental observations of analogue Hawking Radiation, one in general has to keep at least a spatial dimension \ti{free} in order to allow for source/sink of the flow. In such lower dimensional observations at large length scales, anomaly can arise due to the existence of these massive fields at a length scale larger than $1/\xi$ due to the presence of quantum effects at the length scale of order $\xi$. In this paper, we systematically address this issue within a simple mathematical framework.

\section{2. Gross-Pitaevskii (GP) model for Minimal `\ti{Non-local}' $\tb{\ti{s}}$-wave scattering}
A nonuniform BEC is characterised by the \textcolor{black}{general} mean field Gross-Pitaevskii (GP) equation of the form\footnote{This form is derived from Heisenberg representation, see eqn.(5.1) of ref.\cite{Pitaevskii2003}.}
\bea\nonumber 
 && i\hbar\frac{\partial}{\partial t}\psi(\tb{r},t) = \left ( -\frac{\hbar^{2}\na^{2}}{2m} + V_{ext}(\tb{r},t) \right )\psi(\tb{r},t) \\ \label{eq:01}
 &+& \left (\int {d \tb{r}^{\prime}\psi^{*}(\tb{r}^{\prime},t)V(\tb{r}^{\prime} - \tb{r})\psi(\tb{r}^{\prime},t)} \right )\psi(\tb{r},t),
\eea
where $\psi(\tb{r},t)$ is the order parameter and $|\psi(\tb{r},t)|^{2}=n(\tb{r},t)$ is the density of the condensate. 
\par
In the above equation, $m$ is the mass of a boson, \textcolor{black}{$\hbar=h/2\pi$ with $h$ being the Planck constant}, $V_{ext}(\tb{r},t)$ is an external potential and $V(\tb{r}^{\prime}-\tb{r})$ is the interaction potential. \textcolor{black}{Since only the $s$-wave scattering is considered here, at the \ti{diluteness} limit while the $s$-wave scattering length of the system being much smaller than the average inter-particle separation; i.e. $|a| << n^{-1/3}$, one can easily exploit the benefit of substituting the actual interaction potential $V(\tb{r}^{\prime}-\tb{r})$ by somewhat soft effective flat potential $V_{eff}(\tb{r}^{\prime}-\tb{r})$.}
\par
\textcolor{black}{In the usual \ti{local} picture, one makes a 'drastic' delta-function approximation to the range of $V(\tb{r}^{\prime}-\tb{r})$ by considering $\int {d \tb{r}^{\prime}\psi^{*}(\tb{r}^{\prime},t)V(\tb{r}^{\prime} - \tb{r})\psi(\tb{r}^{\prime},t)} \equiv \int {d \tb{r}^{\prime}|\psi(\tb{r}^{\prime},t)|^{2} V_{eff}(\tb{r}^{\prime} - \tb{r})} = \int d\tb{r}^{\prime} |\psi(\tb{r}^{\prime},t)|^{2} g \delta(\tb{r}^{\prime}-\tb{r}) = g |\psi(\tb{r},t)|^{2}$} and writes the local GP equation in the following form
\be \label{eq:02}
i\hbar\frac{\partial}{\partial t}\psi(\tb{r},t) = \left ( -\frac{\hbar^{2}\na^{2}}{2m} + V_{ext}(\tb{r},t) + g|\psi(\tb{r},t)|^{2}  \right )\psi(\tb{r},t),
\ee
 where $g=4\pi\hbar^{2} a/m$ parameterises the strength of the $s$-wave scattering (between different bosons in the gas) considered at the lowest order Born approximation.
\par
Due to the possibility of increasing the $s$-wave scattering length practically from $-\infty$ to $\infty$ near a Feshbach resonance, which has already been experimentally achieved \cite{Cornish2000}, \textcolor{black}{we can take \ti{non-local} $s$-wave scattering into account by considering the interaction potential of the form\footnote{\textcolor{black}{In Eq.\eqref{eq:03}, $\aleph$ is the dimension of the space in which the interactions are considered. Evidently, throughout our analysis, $\aleph = 3$.}}}
\be \label{eq:03}
\textcolor{black}{V_{eff}(\tb{r}^{\prime}-\tb{r}) = \frac{g}{(\sqrt{2\pi}a)^{\aleph}}e^{-\frac{|\tb{r}^{\prime}-\tb{r}|^{2}}{2a^{2}}}}
\ee
and write the GP equation to the leading order approximation in a Taylor expansion of the order parameter in the following form \textcolor{black}{(for detailed analysis, see ref.\cite{Abhijit2015})}
\bea\nonumber 
&\textcolor{black}{i\hbar\frac{\partial}{\partial t}\psi(\tb{r},t) = \left( -\frac{\hbar^{2}}{2m}\na^{2} + V_{ext}(\tb{r},t) + g|\psi(\tb{r},t)|^{2}\right) \psi(\tb{r},t)}& \\ \label{eq:04}
&\textcolor{black}{+ \frac{1}{2} a^{2}g\psi(\tb{r},t)\na^{2}|\psi(\tb{r},t)|^{2} .}&
\eea
The above equation is written in 3D cartesian coordinate under a specific (given by Eq.\eqref{eq:03}) spherical symmetry of interactions.

A change of symmetry might change the small numerical pre-factor of the last term on r.h.s. in Eq.\eqref{eq:04} without affecting the dependence on scattering length $a$ whose critical value would be determined by the density $n$ which is in fact a very large number (usually $\sim 10^{21}$ m.$^{-3}$). Therefore, this numerical pre-factor is not that important.
\par
Note that, even when Feshbach resonance is not there to increase $a$, still to look at the dynamics at smaller length scales one must consider the first minimal correction term to the standard $\delta$-interaction configuration. This is more so given the fact that there already exists a Laplacian in the kinetic term of the dynamics. The detailed justification of considering the non-locality was given in our previous paper \cite{Sarkar2014}.
\par
In Eq.\eqref{eq:04} we are merely considering the lowest order $s$-wave scattering as is taken into account in writing Eq.\eqref{eq:02} with the exception that we are removing the approximation of the $\delta$-correlated particle interactions. \textcolor{black}{The last term $\Big(\frac{1}{2} a^{2}g\psi(\tb{r},t)\na^{2}|\psi(\tb{r},t)|^{2}\Big)$ represents the non-local correction to the local GP equation (i.e. Eq.\eqref{eq:02}) \ti{at its minimal level}. }

\par
Let us note a few features of Eq.\eqref{eq:04} in comparison with the local GP equation (Eq.\eqref{eq:02}). The continuity equation is preserved for this new modified model giving a conservation of mass as that for local GP equation. The ground state solution $\psi_{0}(\tb{r},t)=\sqrt{n(\tb{r},t)}e^{-i\mu t/\hbar}$ of the local GP equation is also a solution of a condensate of Eq.\eqref{eq:04}, where $\mu$ is the chemical potential and the total number of bosons $N=\int |\psi_{0}|^{2} d\tb{r}$ is fixed by the normalisation. The condition of dynamical stability of this so-called ground state will remain the same as the local GP dynamics. 
\par
An important class of solutions of the GP equation is investigated through the \ti{small amplitude excitations} to the ground state $\psi_{0}$.
The order parameter is perturbed around the ground state as
\be \label{eq:05}
\textcolor{black}{\psi(\tb{r},t)=\psi_{0}(\tb{r},t)+\sum_{j}\Big[{u_{j}(\tb{r})e^{-i\omega_{j} t}+v_{j}^{*}(\tb{r})e^{i\omega_{j} t}}\Big]e^{-i\mu t/\hbar} }\hspace{0.1cm},
\ee
where \textcolor{black}{$u_{j}$, $v_{j}$ are the Bogoliubov coefficients} and $\omega_{j}$ is the corresponding excitation frequency of the $j$-th mode.
\par
In order to look at the dispersion relation, a uniform (for simplicity) condensate is considered, \textcolor{black}{i.e. $n$ becomes a real constant}. Now in absence of any external potential, and by taking $\mu=gn$, the linearised dynamics of the small amplitudes emerges as
\bea\nonumber \label{eq:06}
\hbar\omega_{j} u_{j} &=& gnu_{j} + \Big(\textcolor{black}{\frac{a^{2}gn}{2}}-\frac{\hbar^{2}}{2m}\Big)u_{j}^{\prime\prime} + gnv_{j} + \textcolor{black}{\frac{a^{2}gn}{2}}v_{j}^{\prime\prime}, \\\nonumber
-\hbar\omega_{j} v_{j} &=& gnv_{j} + \Big(\textcolor{black}{\frac{a^{2}gn}{2}}-\frac{\hbar^{2}}{2m}\Big)v_{j}^{\prime\prime}+ gnu_{j} + \textcolor{black}{\frac{a^{2}gn}{2}}u_{j}^{\prime\prime}, \\
\eea
with a dispersion relation (considering \textcolor{black}{$u_{j}=u \hspace{0.05cm}e^{i\tb{k}.\tb{r}}$} and \textcolor{black}{$v_{j}=v \hspace{0.05cm}e^{i\tb{k}.\tb{r}}$})
\be \label{eq:07}
\hbar^{2}\omega^{2}=\frac{\hbar^{2}k^{2}gn}{m}+\left (\frac{\hbar^{4}}{4m^{2}} - \textcolor{black}{\frac{\hbar^{2}a^{2}gn}{2m} } \right )k^{4},
\ee
where the 'double-prime' obviously indicates the second-ordered spatial derivative.
\par
 In the absence of $k^{4}$-term in the dispersion relation, the excitations are basically massless phonons with their speed $c_{s}=\sqrt{gn/m}$ in the medium. Corresponding to the model in Eq.\eqref{eq:02}, the healing length is 
 \be \label{eq:08}
 \xi_{0}=\frac{\hbar}{\sqrt{2mgn}}=\frac{1}{\sqrt{8\pi an}} \hspace{0.5cm};
 \ee
 which is considered to be very small compared to the relevant length scales of the dynamics in the usual picture where the quartic term (quantum potential) resulting in the dispersion (Eq.\eqref{eq:07}) is thrown away. 
 \par
 In our model, the healing length becomes 
 \be \label{eq:09}
 \xi=\xi_{0}\textcolor{black}{\left(\frac{1}{2} -  4\pi a^{3}n \right)^{1/2}} \equiv \xi_{0}\epsilon \hspace{0.5cm};
 \ee
  which is tunable by tuning $a$ and can be brought to zero at a rate much faster than $\frac{1}{\sqrt{a}}$ if the stability of the actual BEC permits, see . Independent of the extent of tunability, the correction term and the resulting modified $\xi$ are relevant as one looks at smaller length scales. Note that, shrinking of $\xi$ corresponds to an increase of effective mass.
\par
\textcolor{black}{In our following analysis, we would consider the model in Eq.\eqref{eq:04} instead of the most simplified contact interaction model in Eq.\eqref{eq:02}. However, the main results would equally be valid for the local GP equation (i.e. Eq.\eqref{eq:02}) as well where $\xi$ gets replaced by $\xi_{0}$. So, the following analysis is not at all contingent on adding the additional correction term representing non-locality, but the presence of this correction term makes the analysis more general showing the possible role of tuning the s-wave scattering length $a$ on the dispersion at the level of first minimal approximation.}

\section{3. Emergent gravity} 
Considering a general single particle state of the BEC given by $\psi(\tb{r},t)=\sqrt{n(\tb{r},t)}e^{i\theta(\tb{r},t)/\hbar}$, \textcolor{black}{Eq.\eqref{eq:04} gives rise to a set of coupled equations}.
\par
Now one adds perturbations to the density $n \rightarrow \textcolor{black}{n_{0}} + n_{1}$ and phase $\theta \rightarrow \textcolor{black}{\theta_{0}} + \theta_{1}$ to get the linearised{\footnote{$\textcolor{black}{n_{0}}$ and $\textcolor{black}{\theta_{0}}$ are basically the classical mean-field density and phase respectively. }}  dynamics in $n_{1}$ and $\theta_{1}$ as 
\bea \label{eq:10}
\partial_{t} \hspace{0.05cm}n_{1}+\frac{1}{m}\na . (n_{1}\na \textcolor{black}{\theta_{0}} + \textcolor{black}{n_{0}}\na\theta_{1})=0,\\ \label{eq:11}
\partial_{t} \hspace{0.05cm}\theta_{1} + \frac{\na\textcolor{black}{\theta_{0}} . \na\theta_{1}}{m} + g^{\prime}n_{1} - \frac{\hbar^{2}}{2m}D_{2}n_{1}=0.
\eea
In the Eq.\eqref{eq:11} above, and from now on, the $s$-wave coupling strength $g$ (appearing in Eq.\eqref{eq:02}) has just been replaced by $g^{\prime}$ to avoid a possible clash of notations later. The operator $D_2$ is given by 
\bea\nonumber 
&D_{2}n_{1} = -\frac{n_{1}}{2}\textcolor{black}{n_{0}}^{-3/2}\na^{2}\textcolor{black}{n_{0}}^{1/2}+\frac{\textcolor{black}{n_{0}}^{-1/2}}{2}\na^{2}{(\textcolor{black}{n_{0}}^{-1/2}n_{1})}& \\  \label{eq:12}
&\textcolor{black}{-\frac{g^{\prime} ma^{2}}{\hbar^{2}}\na^{2}n_{1}} \hspace{0.1cm},&
\eea 
where the last term is due to our non-local correction to the local GP equation. Considering a uniform background density of the system (i.e. $\textcolor{black}{n_{0}}$ is a constant) one can write this operator as
\be \label{eq:13}
D_{2} = \frac{2m}{\hbar^{2}}\left( \frac{\hbar^{2}}{4m\textcolor{black}{n_{0}}}-\textcolor{black}{\frac{a^{2}g^{\prime}}{2}}\right ) \na^{2} = \textcolor{black}{\frac{2mg^{\prime}}{\hbar^{2}}\xi^{2}\na^{2}}. 
\ee
When non-local interactions are not taken into account even at the lowest order as has been considered here by us, the second term inside the bracket in the above expression of $D_{2}$ in Eq.\eqref{eq:13} would have been missing.
Now expressing  $n_{1}$ in terms of $\theta_{1}$ puts the continuity equation (i.e. Eq.\eqref{eq:10}) to the following form
\be \label{eq:14}
\partial_{\mu}f_{0}^{\mu \nu}\partial_{\nu}\theta_{1} = 0.
\ee
The matrix elements $f_{0}^{\mu \nu}$ are given by
\bea \nonumber
f_{0}^{00} &=& -(g^{\prime} -\frac{\hbar^{2}}{2m}D_{2})^{-1} ,\\ \nonumber
f_{0}^{0j} &=& -(g^{\prime} -\frac{\hbar^{2}}{2m}D_{2})^{-1}v^{j} ,\\ \nonumber
f_{0}^{i0} &=& -v^{i}(g^{\prime} -\frac{\hbar^{2}}{2m}D_{2})^{-1} ,\\ \label{eq:15}
f_{0}^{ij} &=& \frac{\textcolor{black}{n_{0}}\delta^{ij}}{m}-v^{i}(g^{\prime} -\frac{\hbar^{2}}{2m}D_{2})^{-1}v^{j};
\eea
where $v^{i}=\na^{i}\textcolor{black}{\theta_{0}} / m$ \hspace{0.1cm}are the background velocity field components (Greek indices run from 0 - 3, while Roman indices run from 1 - 3). 
\par
In the standard practice, within the Thomas-Fermi limit, one neglects the terms containing $D_{2}$ considering the large scale variation of density (equivalently phase) fluctuations. Note that, the coefficient of the $D_{2}$ is not actually a smaller term than $g^{\prime}$ because $g^{\prime} = 4\pi \hbar^{2} a/m$. In the absence of non-local interactions, consider the common expression in all the $f_{0}^{\mu\nu}$ in its inverse form as 
\be \label{eq:16}
g^{\prime} - \frac{\hbar^{2}}{4m\textcolor{black}{n_{0}}}\na^{2} = \frac{\hbar^{2}}{2m\textcolor{black}{n_{0}}}\left(\frac{1}{\xi_{0}^{2}}-\frac{1}{2}\na^{2}\right).
\ee
It's obvious from the above equation that, to be able to drop the second term on r.h.s compared to the first in the above expression, the length scale of variation must be much larger than $\xi_{0}$. Starting from the defining inequality of the local GP dynamics $|a| << \textcolor{black}{n_{0}}^{-1/3}$ (condition of diluteness), one can show that $\xi_{0} >> \textcolor{black}{n_{0}}^{-1/3}$. The length scale under consideration, therefore, is effectively much larger than the average inter-particle separations. 
\par
On the basis of this approximation, $f_{0}^{\mu \nu} \rightarrow f^{\mu \nu}$ by dropping the quantum potential term and one casts Eq.\eqref{eq:14} in the standard covariant form
\be \label{eq:17}
\frac{1}{\sqrt{-g}}\partial_{\mu}(\sqrt{-g}g^{\mu \nu}\partial_{\nu})\theta_{1} = 0,
\ee
by identifying $f^{\mu \nu} = \sqrt{-g}g^{\mu \nu}$ with $g=det[g_{\mu \nu}]$ (by definition) and this above Eq.\eqref{eq:17} is the free Klein-Gordon equation for a minimally coupled massless scalar field propagating in a spacetime with inverse metric $g^{\mu \nu}$.

\section{4. The Model}
Throwing out the $D_{2}$ term costs heavily in terms of obscuring the small scale features. To avoid that, one must keep the term containing $D_{2}$ in the full analysis. While doing that we will, from now on, keep the full $D_{2}$ term as is given by Eq.\eqref{eq:13} by taking minimal non-local correction into account. As a result, the healing length under considerations is not $\xi_{0}$ anymore, rather $\xi$ (see Eq.\eqref{eq:09}).
\par
Considering the tunability of $\xi$, at small $\xi$, the elements of $[f_{0}^{\mu \nu}]$ (see Eq.\eqref{eq:15}) can be written to the leading order in $\xi^{2}$ as 
\bea \nonumber
f_{0}^{00} &\textcolor{black}{\simeq}&-\frac{1}{g^{\prime}}(1 + \xi^{2}\na^{2}) ,\\ \nonumber
f_{0}^{0j} &\textcolor{black}{\simeq}& -\frac{1}{g^{\prime}}(1 + \xi^{2}\na^{2})v^{j} ,\\ \nonumber
f_{0}^{i0} &\textcolor{black}{\simeq}& -v^{i}\frac{1}{g^{\prime}}(1 + \xi^{2}\na^{2}) ,\\ \label{eq:18}
f_{0}^{ij} &\textcolor{black}{\simeq}& \frac{\textcolor{black}{n_{0}}\delta^{ij}}{m}-v^{i}\frac{1}{g^{\prime}}(1 + \xi^{2}\na^{2})v^{j}.
\eea
This controlled expansion helps us recover Eq.\eqref{eq:17} as it is as the $\mathcal{O}(1)$ dynamics which would break down at a length scale comparable to the healing length. 
\par
Once we separate the dynamics on independent multiple scales considering 
\be \label{eq:19}
\partial_{\mu} \rightarrow \partial_{\mu} + \epsilon \hspace{0.05cm} \partial_{\mu}^{\prime} \hspace{0.2cm},
\ee  
obviously the order $\xi^{2}$-term (where $\xi^{2}=\xi_{0}^{2}\epsilon^{2}$ with $\epsilon = \textcolor{black}{\sqrt{\frac{1}{2} -4\pi a^{3} n_{0} }}$ being a small parameter; see Eq.\eqref{eq:09}) will act as the source term to the ensuing large scale dynamics. Here, the primed scale is $\epsilon^{-1}$ times larger than the scale at which one gets the Klein-Gordon equation at the intermediate scales. We should be exploiting here this UV-IR coupling to look at the effects at length scales $1/\xi$ and above due to causes at the length scales of 
$\xi$ .
\par
In order to properly separate out the extra $\xi_{0}^{2}\na^{2}$ factor from each element \textcolor{black}{(see Eq.\eqref{eq:18})} of the original metric $[f_{0}^{\mu \nu}]$, we consider the following decomposition 
\be \label{eq:20}
f_{0}^{\mu \nu} \rightarrow f^{\mu \nu} + \epsilon^{2} F^{\mu \nu},
\ee
where this $[F^{\mu \nu}]$ should include the extra $\xi_{0}^{2}\na^{2}$ factor in each of its elements. 
\par
Let us go by this choice of scales as suggested by the dynamics itself with the field expressed in a product form as 
\be  \label{eq:21}
\theta_{1} \rightarrow \theta_{1}^{\prime}(R)\theta_{1}(r),
\ee
where the $\theta_{1}^{\prime}(R)$ varies at larger length scales {\footnote{$R \equiv R^{\mu}$ and $r \equiv r^{\mu}$ are four-vectors over large and small scales respectively.}}; now using Eq.\eqref{eq:19}, Eq.\eqref{eq:20} and Eq.\eqref{eq:21} we get the dynamics for $\theta_{1}^{\prime}(R)$ from Eq.\eqref{eq:14} at $\mathcal{O}(\epsilon)$ as
\bea \nonumber 
&&\Big(\partial_{\mu} f^{\mu \nu}\theta_{1}(r)\Big)\partial_{\nu}^{\prime} \theta_{1}^{\prime}(R) + \partial_{\mu}^{\prime} \theta_{1}^{\prime}(R)\Big(f^{\mu \nu}\partial_{\nu} \theta_{1}(r)\Big)  =0,\\ \label{eq:22}
\\ \nonumber 
&& \mbox{and at $\mathcal{O}(\epsilon^{2})$ as} 
\\ \label{eq:23}
&& \partial_{\mu}^{\prime} f^{\mu\nu}\partial_{\nu} ^{\prime} \theta_{1}^{\prime}(R) + \frac{\partial_{\mu} F^{\mu \nu}\partial_{\nu} \theta_{1}(r)}{\theta_{1}(r)}\theta_{1}^{\prime}(R) = 0. 
\eea
Eq.\eqref{eq:23}, in its structure, is the free Klein-Gordon equation for a \ti{massive} scalar field where the mass term is a function of smaller length scales (hence treated as a constant \textcolor{black}{while seeing the dynamics of $\theta_{1}^{\prime}(R)$}). Eq.\eqref{eq:22} is the \ti{constraint} that has to be obeyed. 

\subsection{\tu{Simple metric}}
Let's consider a general background velocity field $\tb{v}(\tb{r})=\Big(v_{x}(\tb{r})\,, v_{y}(\tb{r})\,, v_{z}(\tb{r})\Big)$ in 3D Cartesian coordinate. 
\par
So, $[f^{\mu\nu}]$ is given by the following form
\be \label{eq:24}
 [f^{\mu\nu}] = \frac{1}{g^{\prime}} \left( \begin{array}{cccc}
-1      &  -v_{x}                 & -v_{y}             & -v_{z}                    
\\
\\
 -v_{x}    & (c_{s}^{2}-v_{x}^{2})                 & -v_{x}v_{y}             & -v_{x}v_{z}                     
\\
\\
-v_{y}       & -v_{y}v_{x}                      & (c_{s}^{2}-v_{y}^{2})            & -v_{y}v_{z}  
\\ 
\\  
-v_{z}      & -v_{z}v_{x}                   & -v_{z}v_{y}            & (c_{s}^{2}-v_{z}^{2})                 
\end{array} \right),
\ee
where 
\be \label{eq:25}
c_{s}^{2} = \textcolor{black}{n_{0}}g^{\prime} / m
\ee 
Well this $c_{s}$ was already introduced towards the end of \ti{Section 2}. ; clearly,
\be \label{eq:26}
det[f^{\mu \nu}] = -\frac{c_{s}^{6}}{g^{\prime 4}}.
\ee
Since, we had already identified $f^{\mu \nu} = \sqrt{-g}g^{\mu \nu}$ (see Eq.\eqref{eq:17} ), so 
\bea \nonumber
det[f^{\mu \nu}] = det[\sqrt{-g}g^{\mu \nu}] = (\sqrt{-g})^{4} \hspace{0.05cm} det[g^{\mu \nu}] \\ \label{eq:27}
= (\sqrt{-g})^{4} \hspace{0.05cm} g^{-1} = g.
\eea
Thus we write the inverse metric as
\be \label{eq:28}
 [g^{\mu\nu}] = \frac{g^{\prime}}{c_{s}^{3}} \left( \begin{array}{cccc}
-1      &  -v_{x}                 & -v_{y}             & -v_{z}                    
\\
\\
 -v_{x}    & (c_{s}^{2}-v_{x}^{2})                 & -v_{x}v_{y}             & -v_{x}v_{z}                     
\\
\\
-v_{y}       & -v_{y}v_{x}                      & (c_{s}^{2}-v_{y}^{2})            & -v_{y}v_{z}  
\\ 
\\  
-v_{z}      & -v_{z}v_{x}                   & -v_{z}v_{y}            & (c_{s}^{2}-v_{z}^{2})                 
\end{array} \right),
\ee
and the effective metric as
\be \label{eq:29}
 [g_{\mu\nu}] = \frac{n}{mc_{s}} \left( \begin{array}{cccc}
-(c_{s}^{2}-\tb{v}^{2})      &  -v_{x}                 & -v_{y}             & -v_{z}                    
\\
\\
 -v_{x}    & 1               & 0           & 0                     
\\
\\
-v_{y}       & 0                      & 1           & 0  
\\ 
\\  
-v_{z}      & 0                  & 0           & 1              
\end{array} \right).
\ee
It's trivial to check $det[g_{\mu \nu}] = g = -c_{s}^{6}/g^{\prime 4}$ which agrees well with both Eq.\eqref{eq:26} and Eq.\eqref{eq:27} simultaneously.
\par
Now the matrix representing the 'source-term' is of the following form
\be \label{eq:30}
[F^{\mu\nu}] = \frac{\xi_{0}^{2}}{g^{\prime}} \left( \begin{array}{cccc}
-\na^{2}             &  -\na^{2}v_{x}                    & -\na^{2}v_{y}              & -\na^{2}v_{z}                  
\\
\\
-v_{x} \na^{2}      & -v_{x}\na^{2}v_{x}              & -v_{x}\na^{2}v_{y}             &  -v_{x} \na^{2} v_{z}                   
\\
\\
-v_{y} \na^{2}                           & -v_{y} \na^{2} v_{x}                                           & -v_{y}\na^{2} v_{y}            & -v_{y} \na^{2} v_{z}
\\ 
\\  
-v_{z} \na^{2}                             & -v_{z} \na^{2} v_{x}                                            & -v_{z} \na^{2} v_{y}             & -v_{z}\na^{2} v_{z}              
\end{array} \right).
\ee

\section{5. A special case: $\tb{An}$ $\tb{example}$}
\ti{Ansatz}- Let's consider the phase-field be Fourier expanded in the modes
\begin{align*}
\theta_{1}(r^{\mu}) \rightarrow Ae^{i(\tb{k}.\tb{r} - \omega t)} \hspace{0.3cm},\hspace{0.3cm}  & \theta_{1}^{\prime}(R^{\mu}) \rightarrow A^{\prime}e^{i(\tb{K}.\tb{R} - W\tau)}
\end{align*}
where pretty obviously 
\begin{align*}
\tb{k} &= (k_{1}, k_{2}, k_{3}) \hspace{0.3cm},\hspace{0.3cm}  & \tb{K} &= (K_{1}, K_{2}, K_{3}) \hspace{0.3cm};
\end{align*}
and $\omega$ is the frequency of the basic field $\theta_{1}$ w.r.t the laboratory frame of reference and $W$ is that of the amplitude $\theta_{1}^{\prime}$. $A$ and $A^{\prime}$ are some constant amplitudes. 
\par
Now we rewrite Eq.\eqref{eq:22} and Eq.\eqref{eq:23} as
\bea \label{eq:31}
(\partial_{\mu}^{\prime} \theta_{1}^{\prime}) \Big(  \partial_{\nu} (f^{\nu \mu} \theta_{1}) + f^{\mu \sigma} \partial_{\sigma} \theta_{1} \Big) = 0, \\ \label{eq:32}
\mbox{and} \hspace{1cm} f^{\mu \nu} ( \partial_{\mu}^{\prime} \partial_{\nu}^{\prime} \theta_{1}^{\prime} ) + \frac{\partial_{\alpha} F^{\alpha \beta} \partial_{\beta} \theta_{1}}{\theta_{1}} \theta_{1}^{\prime} = 0
\eea
respectively.
\par
Since the background flow field defines the metric of the spacetime; one can consider, for the sake of simplicity, a constant velocity field $\tb{v}$ , i.e.
\begin{align*}
v_{x} &= V_{1},  & v_{y} &= V_{2},  & v_{z} &= V_{3},
\end{align*}
with all the $V_{1}, V_{2}, V_{3}$ being constants.
\par
Thus $[f^{\mu \nu}]$ (see Eq.\eqref{eq:24}) takes a rather simplified form where the entries are all constants and essentially the spacetime becomes \ti{flat}.
\par
Now from Eq.\eqref{eq:31}, the above mentioned ansatz for $\theta_{1}(r^{\mu})$ and $\theta_{1}^{\prime}(R^{\mu})$ give rise to
\begin{widetext}
\bean
 \frac{2}{g^{\prime}}  \Bigg[ W \big(\omega - \tb{k}.\tb{v} \big) - K_{1} \bigg(\omega V_{1}+k_{1}(c_{s}^{2} -V_{1}^{2})-k_{2}V_{1}V_{2}-k_{3}V_{3}V_{1} \bigg) 
- K_{2} \bigg(\omega V_{2}-k_{1}V_{1}V_{2}+k_{2}(c_{s}^{2} -V_{2}^{2})-k_{3}V_{2}V_{3} \bigg) \\ 
- K_{3} \bigg(\omega V_{3}-k_{1}V_{3}V_{1}-k_{2}V_{2}V_{3})+k_{3}(c_{s}^{2} -V_{3}^{2} \bigg) \Bigg] \theta_{1}^{\prime} \theta_{1} = 0,
\eean
\end{widetext}
\be \label{eq:33}
\Rightarrow  W = \tb{K}.\tb{v} + \frac{c_{s}^{2}\tb{K}.\tb{k}}{\omega - \tb{k}.\tb{v}} \hspace{0.05cm}.
\ee
Similarly $F^{\mu \nu}$ (see Eq.\eqref{eq:30}) also takes a simpler form and with that in hand, one gets the following relation from Eq.\eqref{eq:32}
\begin{widetext}
\ben
\frac{1}{g^{\prime}} \Big[-c_{s}^{2}\tb{K}^{2}+(W-\tb{K}.\tb{v})^{2} \Big] \theta_{1}^{\prime} + \left(-\frac{\xi_{0}^{2}}{g^{\prime} \theta_{1}}\right)\tb{k}^{2} (\omega - \tb{k}.\tb{v})^{2} \theta_{1}^{\prime} \theta_{1} = 0,
\een
\end{widetext}
\be \label{eq:34}
\Rightarrow  W = \tb{K}.\tb{v} \pm \sqrt{ \xi_{0}^{2}\tb{k}^{2} (\omega - \tb{k}.\tb{v})^{2} + c_{s}^{2}\tb{K}^{2}},
\ee
where the mass term is separately read off as 
\begin{align} \label{eq:35}
M_{0} &= \xi_{0} |\tb{k}| \Omega &(\mbox{with} \hspace{0.2cm} |\omega - \tb{k}.\tb{v}| &= \Omega).
\end{align}
Here $\Omega$ is the frequency of the basic field w.r.t the comoving frame of reference. Appearance of $\xi_{0}$ in the above expression of mass ($M_{0}$) makes the role of the quantum potential quite explicit.

\section{6. Analysis}
Eq.\eqref{eq:33} is the dispersion relation which is always there over and above the massive Klein-Gordon equation (i.e. Eq.\eqref{eq:23} or equivalently Eq.\eqref{eq:32}) which itself in turn gives rise to another dispersion relation given by Eq.\eqref{eq:34}
\par
Now having both Eq.\eqref{eq:33} and Eq.\eqref{eq:34} held simultaneously, it's clearly visible that if the magnitude of wave-vector $( |\tb{K}| )$ has to vanish, then $W=0$ which means either $|\tb{k}|=0$ or $\omega=\tb{k}.\tb{v}$. The simultaneous validity of Eq.\eqref{eq:33} and Eq.\eqref{eq:34}
also implies that the $\tb{K}$ modes are always excited for nonzero $\Omega$.
\par
A selection of $\tb{K}$ modes, given a fixed $\tb{k}$, should come via the combination of both the equation of constraint and the massive KG equation, i.e.
\be\label{eq:36}
\frac{c_{s}^{2}\tb{K}.\tb{k}}{\omega - \tb{k}.\tb{v}} = \pm \sqrt{M_{0}^{2}+ c_{s}^{2}\tb{K}^{2}} \hspace{0.5cm}.
\ee
\textcolor{black}{Given an $M_{0} \neq 0$, for a comoving observer, the $\tb{K}$-modes must be excited above a minimum $|\tb{K}|$ threshold. So, the presence of mass here would observationally rule out a flat amplitude of the basic $\theta_{1}$-modes.} One would also see here an anisotropic velocity of $\tb{K}$ excitations as the existence of $\tb{k}$-modes has broken the symmetry. 
\par
Fig.1 shows a plot of $K_{x}$ vs $M_{0}$ (in 1D for simplicity) for \tb{$_{87}$Rb} with $m \approx 1.45\times 10^{-25}$ kg., the $s$-wave scattering length $a\sim 109 a_{0}$ where $a_{0} \sim 0.529 \times 10^{-10}$ m. is the \ti{Bohr radius}; density of the condensate is $\textcolor{black}{n_{0}}\sim 10^{21}$ m.$^{-3}$. This gives the speed of sound in the condensate to be $c_{s}\sim 61.909\times 10^{-4}$ ms.$^{-1}$ \textcolor{black}{and the healing length $\xi \sim 5.86\times 10^{-8}$ m.} . The frequency of $\theta_{1}$ w.r.t the lab-frame is $\omega = c_{s}|\tb{k}|$. So if a $k_{x}\equiv |\tb{k}|$ is specified, \textcolor{black}{$\omega$ gets fixed automatically. $\Omega$ can easily be eliminated through Eq.\eqref{eq:35} and Eq.\eqref{eq:36} and thus $K_{x}$ is evaluated as a function of $M_{0}$ only while $M_{0}$ being run independently from $0$ to $\sim 0.07$ in S.I. units, $K_{x}$ is plotted accordingly for each fixed $k_{x}$.}

\begin{figure}[h]
  \includegraphics[width=9cm]{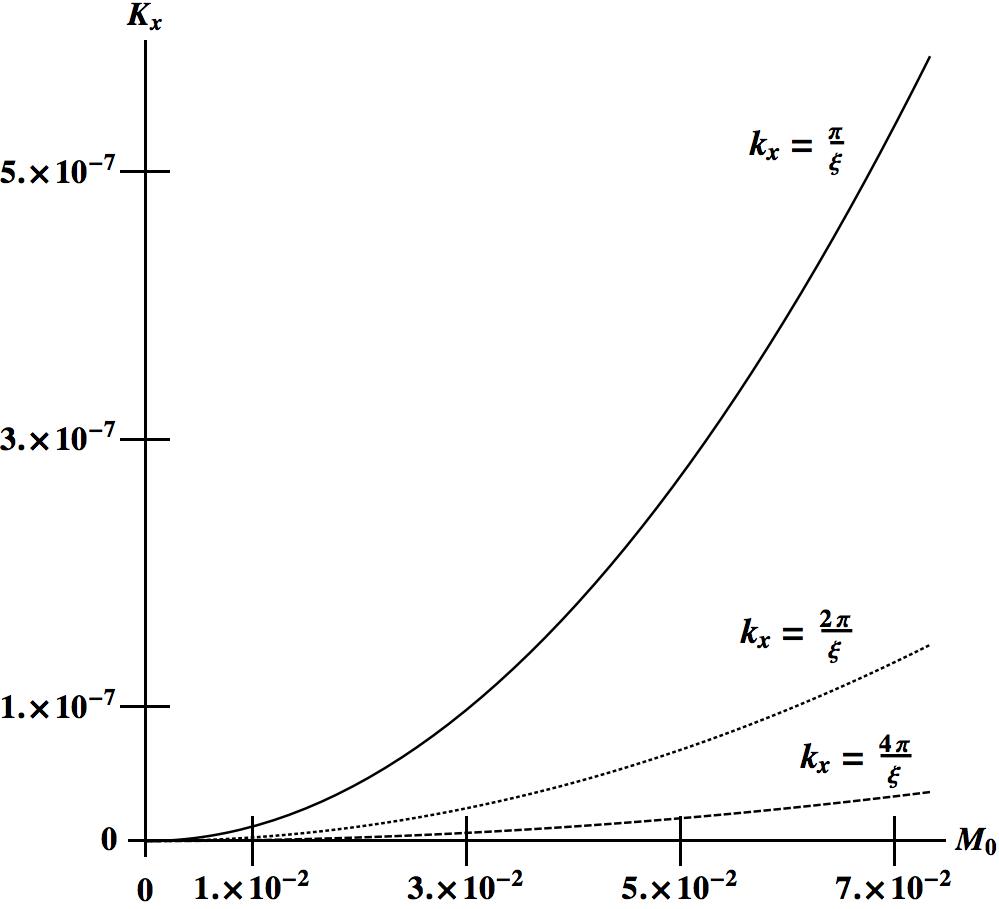}
   \caption {\small {shows the plot of $K_{x}$ vs $M_{0}$ in S.I. units. Here we consider three cases separately for $k_{x}=\frac{\pi}{\xi}, \frac{2\pi}{\xi}, \frac{4\pi}{\xi}$ and they are denoted by continuous, dotted and dashed lines respectively.}}
\end{figure}
\par
A different picture (see Fig.2) arises for an observer on \ti{lower dimensional space}, $x-y$ plane (say). Then the system must be excited at a specified non-zero $K_{z}$ mode at $K_{x}=0=K_{y}$ such that 
\bea \nonumber
\frac{c_{s}^{2}K_{z}k_{z}}{\Omega} = \sqrt{M_{0}^{2}+c_{s}^{2}K_{z}^{2}} \\ \label{eq:37}
\Rightarrow \textcolor{black}{K_{z}= M_{0}^{2} \left( c_{s}^{4}\xi_{0}^{2}|\tb{k}|^{2}k_{z}^{2} - c_{s}^{2}M_{0}^{2} \right)^{-\frac{1}{2}}}.
\eea
This gives the selection of $K_{z}$ which must be excited due to $M_{0} > 0$, or in other words, because of the presence of \ti{quantum potential}.
\textcolor{black}{This situation would be seen by the observer living on $x-y$ plane as the presence of massive phonon-excitations which is supported by the necessary presence of the transverse $K_{z}$-modes. The existence of transverse excitations here are not giving rise to the mass; the mass is entirely determined by the $\tb{k}$-modes in the presence of quantum potential. However, the existence of the transverse modes plays a necessary role in supporting the mass to explicitly show up in lower dimensions. These transverse excitations are unavoidable experimentally and are very important in the sense that they can act as an effective mass for the phonons generated by Hawking emission in an acoustic black hole in a BEC and can cure the infrared divergence which appears in a $(1+1)$ dimensional case \cite{Rinaldi2011,Rinaldi2013}.}

\begin{figure}[h]
  \includegraphics[width=9cm]{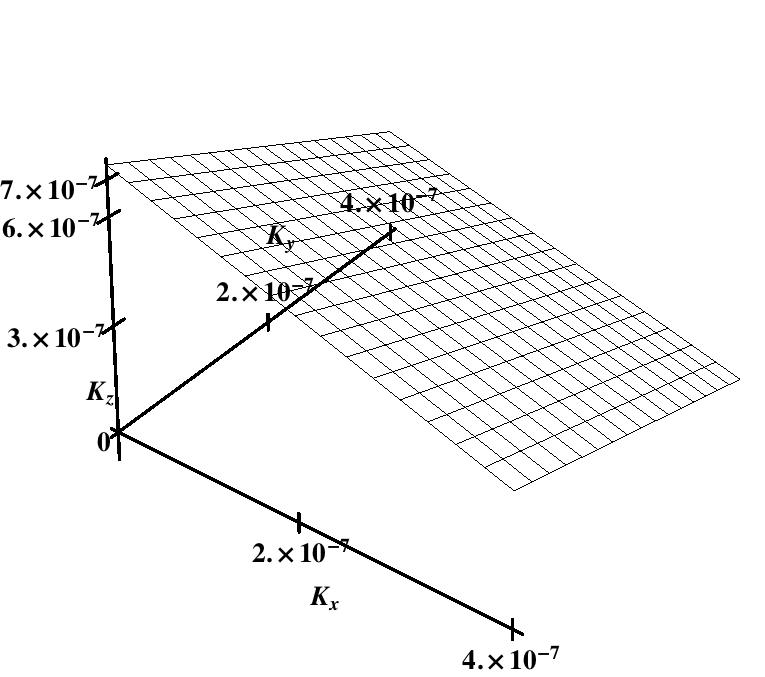}
   \caption {\small {shows the 3D-plot of $K_{z}$ vs $K_{x}-K_{y}$ for $_{87}$\tb{Rb} with $M_{0}=0.25$ in S.I. units, where we've considered $k_{x}=\frac{1}{\surd 3}.\frac{4\pi}{\xi}=k_{y}=k_{z}$ and $Cos(\tb{k},\tb{v})=1$. \textcolor{black}{The planar surface intersects the $K_{z}$-axis at $K_{z}\simeq 7.39\times 10^{-7}$ in S.I. units which agrees with Eq.\eqref{eq:37}.}}}
\end{figure} 
\par
Now for the appearance of $K_{x}$, $K_{y}$ modes, this $K_{z}$ mode can be adjusted such that Eq.\eqref{eq:33} and Eq.\eqref{eq:34} satisfy simultaneously. \textcolor{black}{We have plotted Fig.2 where the regular interdependence between the observed $K_{x}$, $K_{y}$ -modes on $x-y$ plane and the unobserved transverse $K_{z}$ -mode for a given $M_{0}$ is shown. This figure basically shows a wide range of validity of the dispersion relation (i.e. Eq.\eqref{eq:36}) to create massive modes. }


\par
The present result assumes importance in view of the fact that the actual experimental observations of analogue Hawking radiation is always proposed to be in $(2+1)$ dimensions or even lower dimensional space.

\section{7. Discussions}
By a controlled consideration of the quantum potential term in the hydrodynamics of BEC, which could be experimentally achievable, we see the possible emergence of mass based on how close the small scale dynamics approaches to the analogue Compton wavelength of the system. Eq.\eqref{eq:17} has negligible correction if the length scales of the free scalar field is large enough compared to $\xi$. The correction becomes large at small length scales, but, shows up in the large length scale dynamics through an intrinsic coupling between the small and large scales. 
\par
If one speculates the \ti{emergent gravity} in $(3+1)$ dimensions is due to the presence of a 'fictitious' condensate in $(4+1)$ dimension, then the masses seen in $(3+1)$ dimensions are basically due to the extra modes one has to necessarily excite in $(4+1)$ dimensions. The coupling takes place here through the quantum potential term which does break the Lorentz symmetry. Because of this small scale Lorentz-breaking in higher dimensions, the larger scale dynamics in lower dimensions get affected. 


\par
Apart from the above mentioned fictitious scenario, the present analysis has practical implications in experimentally looking for the Hawking spectrum in condensates. In general, if Hawking radiation to be seen in laboratories, it has to be seen in $(2+1)$ dimensions or even lower dimensional spacetime. This is because one has to keep at least one dimension free to handle the source/sink of the flow. 
\par
Under such considerations, any excitations in the left out dimensions would support massive excitations to appearing through the quantum potential coupling and these excitations could be seen at larger length scales. Proper correction has to be taken into account to filter out the \ti{expected} Hawking spectrum in such a scenario and the framework presented here can be used for that purpose. 
\par
\textcolor{black}{The presence of this quantum potential induced mass would also be felt in the phonons excited in the $(3+1)$ dimensions since one cannot excite a constant finite amplitude wave number (nonzero $|\tb{k}|$) mode. There would necessarily be an energy expense in exciting the associated amplitude modulations for nonzero $\tb{k}$-modes in $(3+1)$ dimensions due to the presence of the quantum potential term.}


\subsection{ Acknowledgement} 
AB acknowledges very useful discussions with \href{http://www.iiserpune.ac.in/~suneeta/}{\tb{Suneeta Vardarajan}}.

\newpage
\section{Appendix}

In this appendix, we provide the detailed derivation of the presence of the correction term on r.h.s. in Eq.\eqref{eq:04} to incorporate non-locality at its minimal level. The actual interaction term ($T_{int}$), from Eq.\eqref{eq:01}, is given by
\be \label{eq:38}
T_{int}= \psi(\tb{r},t) \int_{\tb{r}^{\prime}=\tb{r}}^{\infty} {d \tb{r}^{\prime} \hspace{0.1cm} |\psi(\tb{r}^{\prime},t)|^{2} \hspace{0.05cm}V(\tb{r}^{\prime} - \tb{r})}
\ee
Now we Taylor expand $|\psi(\tb{r}^{\prime},t)|^{2}$ in 3D about the point $\tb{r}^{\prime}=\tb{r}$ given by 
\bea \nonumber
|\psi(\tb{r}^{\prime},t)|^{2} = |\psi(\tb{r},t)|^{2} + (\tb{r}^{\prime}-\tb{r}).\na^{\prime}|\psi(\tb{r}^{\prime},t)|^{2}\Bigg|_{\tb{r}^{\prime}=\tb{r}}& \\ \nonumber
+\frac{1}{2}\Big( (\tb{r}^{\prime}-\tb{r}).\na^{\prime} \Big)^{2}|\psi(\tb{r}^{\prime},t)|^{2}\Bigg|_{\tb{r}^{\prime}=\tb{r}} + ...& \\ \label{eq:39}
\eea
and the interaction potential $V(\tb{r}^{\prime} - \tb{r})$, in the context of $s$-wave scattering, is evidently substituted by some effective soft potential $V_{eff}(\tb{r}^{\prime} - \tb{r}) = \frac{g}{(\sqrt{2\pi}a)^{3}}e^{-\frac{|\tb{r}^{\prime}-\tb{r}|^{2}}{2a^{2}}}$ (see Eq.\eqref{eq:03}).
\par
Hence from Eq.\eqref{eq:38}, we have
\begin{widetext}
\bea \nonumber
T_{int} &=& \psi(\tb{r},t) \int_{\tb{r}^{\prime}=\tb{r}}^{\infty} {d \tb{r}^{\prime} \hspace{0.1cm} \Bigg(|\psi(\tb{r},t)|^{2} + (\tb{r}^{\prime}-\tb{r}).\na^{\prime}|\psi(\tb{r}^{\prime},t)|^{2}\Bigg|_{\tb{r}^{\prime}=\tb{r}}+\frac{1}{2}\Big( (\tb{r}^{\prime}-\tb{r}).\na^{\prime} \Big)^{2}|\psi(\tb{r}^{\prime},t)|^{2}\Bigg|_{\tb{r}^{\prime}=\tb{r}} + ...\Bigg) \hspace{0.05cm}V_{eff}(\tb{r}^{\prime} - \tb{r})} \\ \nonumber
 &=& \psi(\tb{r},t) \int_{\tb{r}^{\prime}=\tb{r}}^{\infty} {d \tb{r}^{\prime} \hspace{0.1cm} \Bigg(|\psi(\tb{r},t)|^{2} + (\tb{r}^{\prime}-\tb{r}).\na^{\prime}|\psi(\tb{r}^{\prime},t)|^{2}\Bigg|_{\tb{r}^{\prime}=\tb{r}}+\frac{1}{2}\Big( (\tb{r}^{\prime}-\tb{r}).\na^{\prime} \Big)^{2}|\psi(\tb{r}^{\prime},t)|^{2}\Bigg|_{\tb{r}^{\prime}=\tb{r}} + ...\Bigg) \hspace{0.05cm}\frac{g}{(\sqrt{2\pi}a)^{3}}e^{-\frac{|\tb{r}^{\prime}-\tb{r}|^{2}}{2a^{2}}}} \\ \label{eq:40}
 &=& T_{int}^{(0)} \hspace{0.5cm}+\hspace{0.5cm} T_{int}^{(1)} \hspace{0.5cm}+\hspace{0.5cm} T_{int}^{(2)} + ... 
\eea
\end{widetext}
where quite obviously, 
\begin{widetext}
\bea \nonumber
T_{int}^{(0)} &=& \psi(\tb{r},t) \int_{\tb{r}^{\prime}=\tb{r}}^{\infty} {d \tb{r}^{\prime} \hspace{0.1cm} |\psi(\tb{r},t)|^{2} \hspace{0.05cm}\frac{g}{(\sqrt{2\pi}a)^{3}}e^{-\frac{|\tb{r}^{\prime}-\tb{r}|^{2}}{2a^{2}}}} \hspace{0.5cm}=\hspace{0.5cm} \psi(\tb{r},t) |\psi(\tb{r},t)|^{2} \frac{g}{(\sqrt{2\pi}a)^{3}} \int_{\tb{r}^{\prime}=\tb{r}}^{\infty} {d \tb{r}^{\prime} \hspace{0.1cm}  e^{-\frac{|\tb{r}^{\prime}-\tb{r}|^{2}}{2a^{2}}}} \\ \nonumber
&=& \psi(\tb{r},t) |\psi(\tb{r},t)|^{2} \frac{g}{(\sqrt{2\pi}a)^{3}} \int_{-\infty}^{\infty} {d x^{\prime}  e^{-\frac{|x^{\prime}-x|^{2}}{2a^{2}}}}\int_{-\infty}^{\infty} {d y^{\prime}  e^{-\frac{|y^{\prime}-y|^{2}}{2a^{2}}}}\int_{-\infty}^{\infty} {d z^{\prime}  e^{-\frac{|z^{\prime}-z|^{2}}{2a^{2}}}}  \\ \nonumber
&=& \psi(\tb{r},t) |\psi(\tb{r},t)|^{2} \frac{g}{(\sqrt{2\pi}a)^{3}} \hspace{0.3cm}\Big(\int_{-\infty}^{\infty} {d x^{\prime}  e^{-\frac{|x^{\prime}-x|^{2}}{2a^{2}}}}\Big)^{3} \hspace{0.2cm}=\hspace{0.2cm}  \psi(\tb{r},t) |\psi(\tb{r},t)|^{2} \frac{g}{(\sqrt{2\pi}a)^{3}} \hspace{0.1cm}(\sqrt{2\pi}a)^{3} \\ \label{eq:41}
&=& g \psi(\tb{r},t) |\psi(\tb{r},t)|^{2} \hspace{0.2cm},
\eea
\end{widetext}
\begin{widetext}
\bea \nonumber
T_{int}^{(1)} &=& \psi(\tb{r},t) \int_{\tb{r}^{\prime}=\tb{r}}^{\infty} {d \tb{r}^{\prime} \hspace{0.1cm}\Bigg((\tb{r}^{\prime}-\tb{r}).\na^{\prime}|\psi(\tb{r}^{\prime},t)|^{2}\Bigg|_{\tb{r}^{\prime}=\tb{r}} \Bigg) \hspace{0.05cm}\frac{g}{(\sqrt{2\pi}a)^{3}}e^{-\frac{|\tb{r}^{\prime}-\tb{r}|^{2}}{2a^{2}}}}  \\ \nonumber
&=& \psi(\tb{r},t) \frac{g}{(\sqrt{2\pi}a)^{3}}  \int_{-\infty}^{\infty} \int_{-\infty}^{\infty} \int_{-\infty}^{\infty} d x^{\prime} \hspace{0.1cm} d y^{\prime} \hspace{0.1cm} d z^{\prime}  \hspace{0.2cm} \Bigg((x^{\prime}-x)\partial_{x^{\prime}}+ (y^{\prime}-y)\partial_{y^{\prime}} + (z^{\prime}-z)\partial_{z^{\prime}} \Bigg)|\psi(\tb{r}^{\prime},t)|^{2}\Bigg|_{\tb{r}^{\prime}=\tb{r}} e^{-\frac{|\tb{r}^{\prime}-\tb{r}|^{2}}{2a^{2}}}  \\ \label{eq:42}
&=& \psi(\tb{r},t) \frac{g}{(\sqrt{2\pi}a)^{3}}  3\Bigg(\partial_{x}|\psi(\tb{r},t)|^{2}\int_{-\infty}^{\infty}d z^{\prime} \int_{-\infty}^{\infty} d y^{\prime} \underbrace{\int_{-\infty}^{\infty} d x^{\prime}   \hspace{0.2cm} (x^{\prime}-x) e^{-\frac{|\tb{r}^{\prime}-\tb{r}|^{2}}{2a^{2}}}}\Bigg) = 0 \hspace{0.2cm},
\eea
the above under-braced part vanishes since the integrand is an odd function while the integration limits are from $-\infty$ to $+\infty$. Its evident that all the odd higher order terms, viz. $T_{int}^{(3)}$, $T_{int}^{(5)}$,... etc. would vanish likewise.
\end{widetext}
\begin{widetext}
Now,
\bea \nonumber
T_{int}^{(2)} &=& \psi(\tb{r},t) \int_{\tb{r}^{\prime}=\tb{r}}^{\infty} {d \tb{r}^{\prime} \hspace{0.1cm}\Bigg(\frac{1}{2}\Big((\tb{r}^{\prime}-\tb{r}).\na^{\prime}\Big)^{2}|\psi(\tb{r}^{\prime},t)|^{2}\Bigg|_{\tb{r}^{\prime}=\tb{r}} \Bigg) \hspace{0.05cm}\frac{g}{(\sqrt{2\pi}a)^{3}}e^{-\frac{|\tb{r}^{\prime}-\tb{r}|^{2}}{2a^{2}}}}  \\ \nonumber
&=& \psi(\tb{r},t) \frac{g}{(\sqrt{2\pi}a)^{3}} \int_{-\infty}^{\infty} \int_{-\infty}^{\infty} \int_{-\infty}^{\infty} d x^{\prime} \hspace{0.1cm} d y^{\prime} \hspace{0.1cm} d z^{\prime}  \hspace{0.2cm} \frac{1}{2}\Bigg((x^{\prime}-x)\partial_{x^{\prime}}+ (y^{\prime}-y)\partial_{y^{\prime}} + (z^{\prime}-z)\partial_{z^{\prime}} \Bigg)^{2}|\psi(\tb{r}^{\prime},t)|^{2}\Bigg|_{\tb{r}^{\prime}=\tb{r}} e^{-\frac{|\tb{r}^{\prime}-\tb{r}|^{2}}{2a^{2}}}  \\ \nonumber
&=& \psi(\tb{r},t) \frac{g}{(\sqrt{2\pi}a)^{3}} \frac{1}{2} \int_{-\infty}^{\infty} \int_{-\infty}^{\infty} \int_{-\infty}^{\infty} d x^{\prime} \hspace{0.1cm} d y^{\prime} \hspace{0.1cm} d z^{\prime}  \hspace{0.2cm} \Bigg((x^{\prime}-x)^{2}\hspace{0.1cm}\partial_{x^{\prime}x^{\prime}}+(y^{\prime}-y)^{2}\hspace{0.1cm}\partial_{y^{\prime}y^{\prime}}+(z^{\prime}-z)^{2}\hspace{0.1cm}\partial_{z^{\prime}z^{\prime}}\Bigg)|\psi(\tb{r}^{\prime},t)|^{2}\Bigg|_{\tb{r}^{\prime}=\tb{r}} e^{-\frac{|\tb{r}^{\prime}-\tb{r}|^{2}}{2a^{2}}} \\ \nonumber
&=& \psi(\tb{r},t) \frac{g}{(\sqrt{2\pi}a)^{3}} \frac{1}{2} \Big(\partial_{xx} + \partial_{yy} + \partial_{zz}\Big)|\psi(\tb{r},t)|^{2} \int_{-\infty}^{\infty} \int_{-\infty}^{\infty} \int_{-\infty}^{\infty} d x^{\prime} \hspace{0.1cm} d y^{\prime} \hspace{0.1cm} d z^{\prime}  \hspace{0.2cm} (x^{\prime}-x)^{2} e^{-\frac{|\tb{r}^{\prime}-\tb{r}|^{2}}{2a^{2}}} \\ \label{eq:43}
&=& \frac{1}{2}a^{2}g\psi(\tb{r},t)\na^{2}|\psi(\tb{r},t)|^{2}\hspace{0.2cm}.
\eea
\end{widetext}
Using Eq.s\eqref{eq:41}, \eqref{eq:42} and \eqref{eq:43}, the full interaction term in Eq.\eqref{eq:40} is given by
\be \label{eq:44}
T_{int} = g \psi(\tb{r},t) |\psi(\tb{r},t)|^{2} + 0 + \frac{1}{2}a^{2}g\psi(\tb{r},t)\na^{2}|\psi(\tb{r},t)|^{2} + ...
\ee
In order to incorporate the non-locality, here we have considered the first minimal correction on top of the local picture and thus the presence of the last term on r.h.s of Eq.\eqref{eq:04} is well justified.
\par
From \ti{Section 3}, we substituted $g$ by $g^{\prime}$ in order to avoid possible clash of notations and the correction term $\frac{1}{2}a^{2}g^{\prime}\psi(\tb{r},t)\na^{2}|\psi(\tb{r},t)|^{2} $ is evidently incorporated in the expression of $D_{2}$, see Eq.\eqref{eq:12}.

\bibliographystyle{apsrev4-1}
\bibliography{Q_potentialFinal_bib}

\end{document}